\documentclass[preprint,aps]{revtex4}
\usepackage{epsfig,amsmath,amssymb}
\newcommand{\be}{\begin{equation}}
\newcommand{\ee}{\end{equation}}

\newcommand{\bit}{\begin{itemize}}
\newcommand{\eit}{\end{itemize}}
\newcommand{\bea}{\begin{eqnarray}}
\newcommand{\eea}{\end{eqnarray}}
\newcommand{\stateX}{\(\sqrt{3}\)\(\times\)\(\sqrt{3}\)}
\newcommand{\stateO}{\(q\)\(=\)\(0\)}

\newcommand{\kagome}{{ kagom\'e }}
\newcommand{\skagome}{{ square-kagom\'e }}
\newcommand{\Kagome}{{\Kagome} }
\newcommand{\ebond}{e_0}
\begin{document}

\title{
The spin-half Heisenberg antiferromagnet on the 
{\em square-kagom\'e}
lattice: Ground state and low-lying excitations 
}

\author{J. Richter}
\affiliation{Institut f\"ur Theoretische Physik, Universit\"at Magdeburg,
      P.O. Box 4120, D-39016 Magdeburg, Germany}
\author{J. Schulenburg}
\affiliation{Universit\"{a}tsrechenzentrum,
             Universit\"{a}t Magdeburg,
             P.O. Box 4120, D-39016 Magdeburg, Germany}
\author{P. Tomczak}
\affiliation
{
Physics Department, Adam Mickiewicz University, 
Umultowska 85, 61-614 Pozna\'n, Poland
}
\author{D. Schmalfu{\ss}}
\affiliation{ Institut f\"ur Theoretische Physik, Universit\"at Magdeburg,
      P.O. Box 4120, D-39016 Magdeburg, Germany}
\date{\today}

\begin{abstract}
We discuss the ground state and the low-lying excitations 
of the spin-half Heisenberg
antiferromagnet on the two-dimensional 
\skagome lattice.
This magnetic system belongs to the class of highly frustrated spin systems 
with
an infinite non-trivial
degeneracy of the classical ground state
as it is known also for the  Heisenberg
antiferromagnet on the \kagome and on the star
lattice.  
The quantum ground state of the spin-half 
system is a quantum paramagnet likely with a finite spin gap and with a  
large number of non-magnetic excitations within this gap.
The magnetization versus field curve
shows 
plateaux as well as a macroscopic magnetization
jump to saturation due to independent localized magnon states.
\end{abstract}


\maketitle

\section{Introduction} 
The magnetic  properties of low-dimensional antiferromagnetic 
quantum spin systems have been a
subject of many theoretical studies in recent years. 
These studies are motivated by 
the recent progress in 
synthesizing quasi-two-dimensional magnetic materials 
which exhibit exciting quantum effects 
\cite{taniguchi95, kageyama99, coldea, TOKMIG02,lemmens,harris}.

A lot of activities in this area
were focused on  frustrated spin-half Heisenberg antiferromagnets
like the
$J_1$-$J_2$ antiferromagnet on the square lattice
(see, e.g. Refs. \cite{squa_ref7,squa_ref8,squa_ref9,
squa_ref10,Sushkov01,squa_ref13a,capriotti01} 
and references therein)
 and on the cubic \cite{schmidt02,oitmaa04} lattice, the 
Heisenberg antiferromagnet (HAFM) 
on the star lattice \cite{RiSch04,Richter04} and last but not least 
the HAFM on the  
\kagome lattice 
(see the reviews [\onlinecite{Richter04,lhuillier03,lhuillier01,moessner01}] and
references therein). 
Due to the extreme frustration the 
HAFM on the \kagome and the star lattices shows an infinite 
non-trivial degeneracy of the
classical ground state. Furthermore, both spin
lattices exhibit a magnetization jump to
saturation due to localized magnon states
\cite{prl02,ri04,RiSch04,Richter04}. 
Although there is most likely no magnetic ground state order 
for the quantum spin-half HAFM on both lattices,
the nature of both quantum ground states and also the low-lying spectrum 
are basically different.
It was argued \cite{RiSch04}
that the origin for this difference lies in the existence of 
non-equivalent nearest-neighbor (NN)
bonds in the star lattice whereas all NN bonds in the \kagome lattice are
equivalent. Another striking difference relevant for magnetic
properties \cite{sachdev} 
lies in the number of spins in the
unit cell which is odd for the \kagome lattice but even for the star lattice. 
As a result of the interplay between quantum fluctuations and
strong frustration  for the  \kagome lattice 
the quantum ground state is  a quantum spin liquid with 
very short-ranged spin, dimer, and chirality correlations (see e.g.
 \cite{lecheminant97,waldtmann98,lhuillier03, Richter04, lhuillier01}),
a (small) spin gap to the 
triplet excitations
and an exceptional density of low-lying singlets below the first magnetic
excitation.
On the other hand, for the star lattice one meets a 
so-called {\it explicit valence-bond crystal} with a well-pronounced gap to
all excitations which can be attributed to the non-equivalence of the NN
bonds and to the even number of $s=1/2$ spins in the unit cell
 \cite{Richter04, RiSch04}. 

\begin{figure}[t!]
\centerline{\epsfig{file=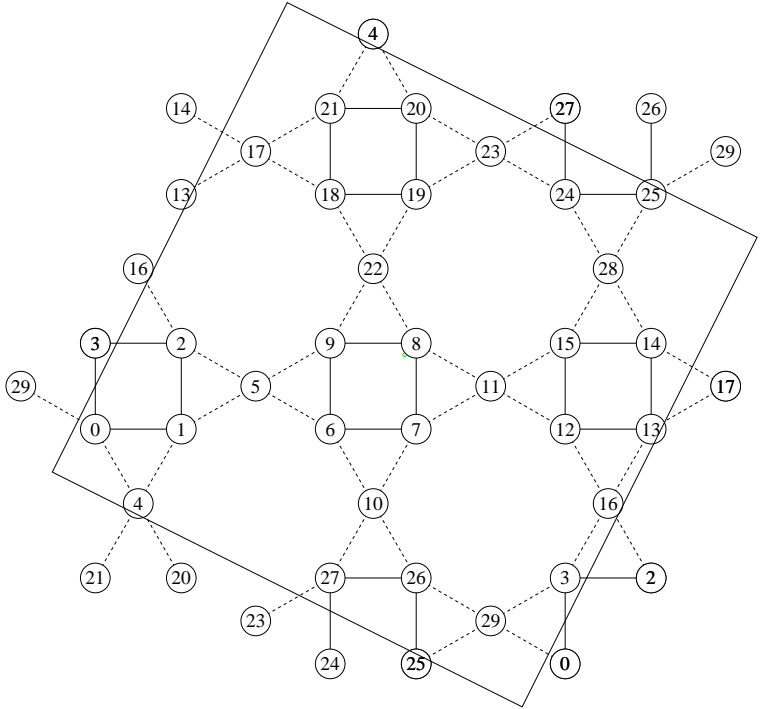,width=10cm}}
\caption[]{
The \skagome lattice with $N=30$ sites. The two topologically inequivalent 
nearest-neihghbor
bonds are distinguished by solid (square bonds $J_S$)
and dashed lines 
(triangular bonds $J_T$).
} 
\label{fig1}
\end{figure}

In this paper we consider the spin-half HAFM 
on the \skagome  \cite{Sidd,tomczak03,sp04}
lattice (see Fig.~\ref{fig1}). 
The \skagome lattice is built  by regular but also by non-regular polygons
and it has two non-equivalent sites. Therefore it does not 
belong to  the class of so-called  uniform tilings \cite{gruenbaum,Richter04}
(like, e.g. square, triangular, star 
or \kagome lattice). Nevertheless, 
there exist some important geometrical similarities 
to the  \kagome and also to the star lattices.
Similar to the \kagome lattice it has coordination number $z=4$, 
 the even regular polygons (hexagons for the kagom\'e,
squares for the \skagome lattice) are surrounded only by odd regular polygons (triangles) 
and both lattices contain corner sharing triangles.
As a result  the HAFM on the \skagome lattice 
is also strongly frustrated and exhibits  
an infinite non-trivial degeneracy of the classical ground state.
The similarity to the star lattice consists in the existence of
non-equivalent NN bonds and in the fact that both lattices  
have an even number of spins in the unit cell. Moreover
the classical ground state of the HAFM
on the star lattice also exhibits an infinite non-trivial degeneracy.
Due to these similarities we  can expect that  
the HAFM on the  \skagome lattice is  
another candidate for a quantum paramagnetic 
ground state. However, 
the question arises, whether the quantum  
ground state displays similar properties as that for the  \kagome lattice or 
as that for
the star lattice or none of them.
%

\section{The Model}
The geometric unit cell of the \skagome lattice contains six
sites and  the underlying Bravais lattice is a square one
(see Fig.~\ref{fig1}).
For this lattice we consider the spin-half HAFM in a  magnetic field $h$
\begin{equation}
\label{Ham1}
 \hat{H} = J \sum_{\langle ij \rangle}{\bf S}_i \cdot {\bf S}_j
- h \hat{S}^z,
\end{equation}
where the sum runs over pairs of neighboring sites $\langle ij \rangle$ and
$\hat{S}^z = \sum_i\hat{S}^z_i$.
As mentioned above the \skagome lattice carries topologically inequivalent 
NN
bonds $J_S$ (square bonds, solid lines in Fig.~\ref{fig1}) and $J_T$
(triangular bonds, dashed lines in Fig.~\ref{fig1}, see also
Fig.~\ref{fig2}).
For the uniform lattice these bonds are of equal strength $J_S=J_T=J$ and we
set  $J=1$ in what follows.

\section {Semi-classical ground state} 
\begin{figure}[t!]
\centerline{\epsfig{file=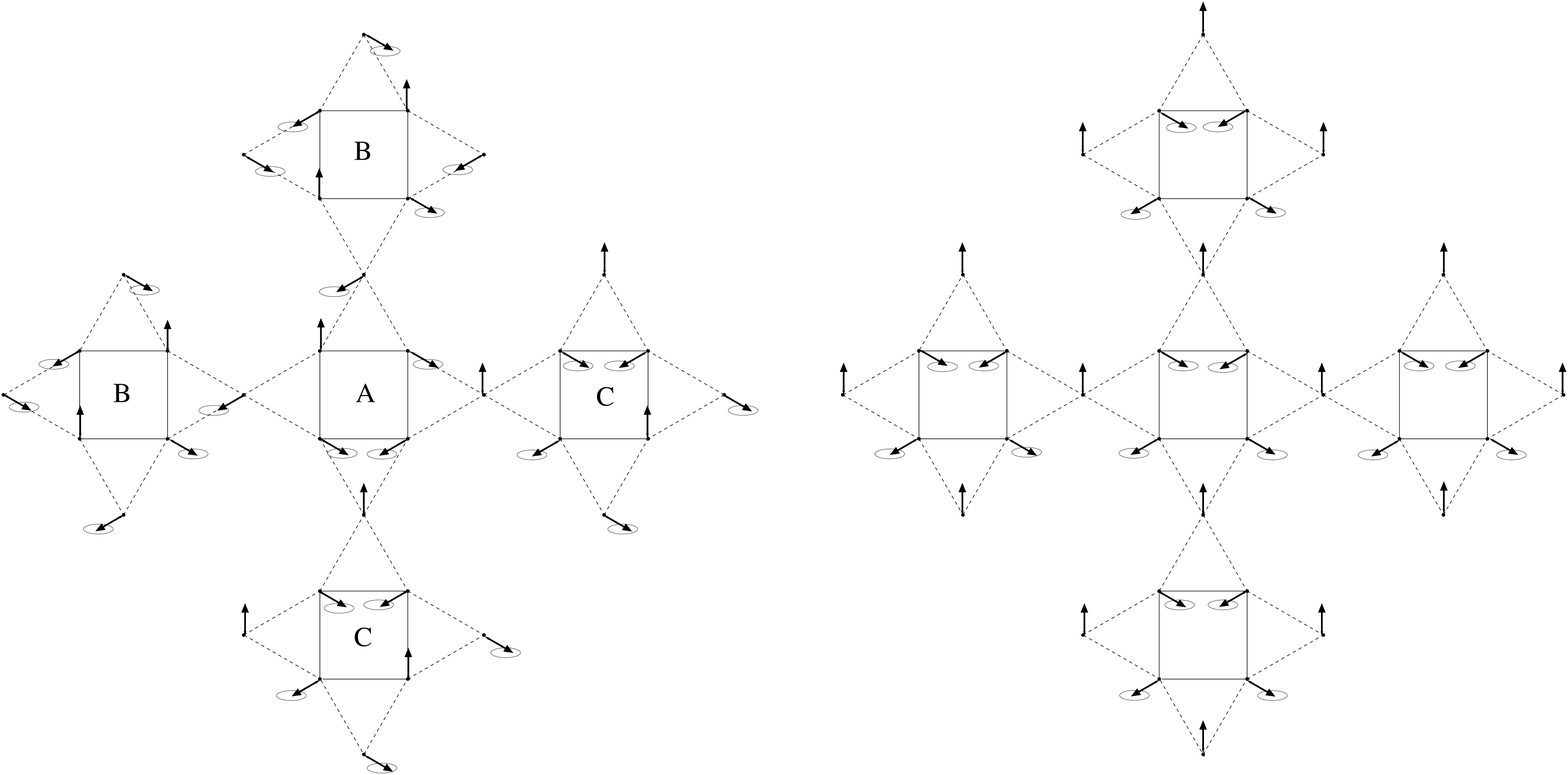,width=1.\columnwidth}}
\caption[]{
Two variants of the ground state of the classical HAFM on the \skagome lattice:
The state on the left side has a magnetic unit cell which is three 
times as large as the geometric one and resembles the \stateX$ $ state of
the \kagome and the star lattices.  For the  state on the right side the
magnetic unit cell is identical to the geometric one and corresponds to the 
 \stateO$ $ state of the \kagome and the star lattices.
The dotted ellipses show further degrees of freedom of the highly
degenerate classical ground state.
}
\label{fig2}
\end{figure}

In the classical ground state for $h=0$ the angle between neighboring spins
is  $2\pi/3$. Since the triangles are "corner sharing",
there is a non-trivial infinite degeneracy
resulting from  the possible rotation of two spins on a triangle (see also
Fig.~\ref{fig2}).
The classical ground state energy per bond is 
$e_0^{\rm class} = -0.125$ assuming classical spin vectors of length $s=1/2$.
Similar to the \kagome and the star lattices there are 
two variants of the  classical ground state,  
shown in Fig.~\ref{fig2}, being candidates for possible magnetic ground state
ordering.

To discuss the influence of quantum fluctuations on a semiclassical level 
we perform a linear spin-wave theory (LSWT)
starting from the coplanar classical ground states.
We have to consider six types of magnons according to the six sites per
unit cell.
As for the kagom\'{e} \cite{harris92,ChaHoShe,asakawa94} and the star
lattice \cite{RiSch04}
the
spin-wave spectra are equivalent for all coplanar 
configurations satisfying the classical ground state constraint.
We obtain six spin-wave branches, three optical branches, one acoustical and  
two dispersionless zero modes.
Thus also flat zero modes appear as it is observed for the \kagome and 
star lattice case.
There is no `order-by-disorder' selection among the coplanar classical
ground states due to the equivalence of the spin-wave branches obtained from
LSWT, exactly like for the \kagome lattice \cite{ChaHoShe,moessner01} and
the star lattice \cite{RiSch04}.

The ground state energy per bond for $s=1/2$ in the LSWT is 
$\ebond = -0.236555$.
Due to the flat zero modes the integral for the sublattice 
magnetization diverges \cite{asakawa94} which might be understood
as some hint for the absence of the classical order. 
Although on the semiclassical LSWT level  the square-kagom\'e, 
the \kagome and the 
star lattices
exhibit almost identical properties, the situation will be changed taking
into account the quantum fluctuations more properly.

\noindent
\section {Exact diagonalization}
To take into account the quantum fluctuations going beyond the
semiclassical LSWT we use  
Lanczos exact diagonalization (ED)  
to calculate the ground state and the lowest
excitations   for the $s=1/2$ HAFM at $h=0$ on finite lattices
of $N=12,18,24,30, 36$  sites with periodic boundary conditions.
The ground states of all those systems are  singlets and the ground state 
energy per
bond $e_0$ and the degeneracy of the quantum ground state $d_{GS}$ 
are  given in 
Table \ref{tab1}.
Furthermore, we give in Table \ref{tab1} the  
gap to the first triplet excitation (spin gap) $\Delta$.
Note that $e_0$ and $\Delta$ are significantly smaller than the
corresponding values  for the star lattice but of comparable size as the
values for the \kagome lattice. 

\begin{figure}
\centerline{\epsfig{file=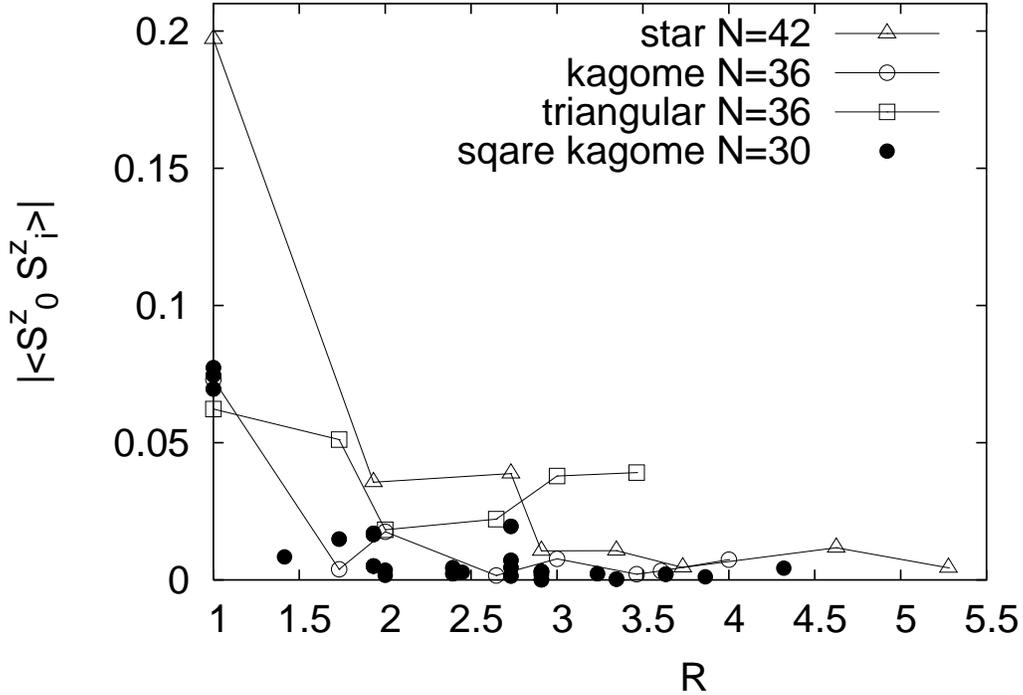,width=0.6\columnwidth, angle=270}}
\caption[]{
The absolute value the spin-spin correlations 
$|\langle {\bf S}_i{\bf S}_j \rangle|$ versus 
$R=|{\bf R}_i - {\bf R}_j|$ for
the HAFM on the \skagome $(N=30)$, the \kagome  $(N=36)$, the star  $(N=42)$
and the triangular $(N=36)$ lattices. For the kagom\'e, star  
and triangular lattices we present only the maximal values of 
$|\langle {\bf S}_i{\bf S}_j \rangle|$ for a certain separation $R$
(the lines are guides for the eyes), 
for the \skagome lattice we present {\bf all} different values for 
$|\langle {\bf S}_i{\bf S}_j \rangle|$ obtained by averaging over the
four degenerate ground states.    
Note that the data on for the
\kagome lattice coincide with those of Ref. \onlinecite{leung93} and  the date for
the triangular and the star lattice with those of Ref.  \onlinecite{RiSch04}.   
} 
\label{sij}
\end{figure}

Now we compare the spin-spin correlations with 
those for the HAFM on the triangular, kagom\'e and star lattices in
Fig~\ref{sij}. 
For the triangular, kagom\'e and star lattices we consider the 
strongest correlations as a measure for
magnetic order for the largest finite lattices accessible for ED
and present in Fig.~\ref{fig2}
the maximal absolute correlations
$|\langle \hat{S}^z_i \hat{S}^z_j\rangle|_{\max}$
for a certain separation $R=|{\bf R}_i - {\bf R}_j|$ versus $R$. 
Contrary to those lattices  the \skagome lattice contains two inequivalent sites.
Hence we  
present for the \skagome lattice {\bf all} different correlations 
$|\langle \hat{S}^z_i \hat{S}^z_j \rangle|$ in Fig~\ref{sij}.
Note further that we prefer to present the correlations for 
the finite \skagome lattice with $N=30$ sites, 
since it has better geometrical
properties than the largest \skagome lattice considered ($N=36$).
As expected we have very rapidly decaying correlations for the disordered 
kagom\'e and star case, whereas the correlations for the N\'{e}el ordered
triangular lattice are much stronger for larger distances and 
show a kind of saturation for larger $R$.
The decay of the correlations for the \skagome lattice is also very rapid
thus indicating the lack of long-range order in the spin-spin
correlation function.
The two non-equivalent NN bonds carry
very similar spin correlations, its difference for $N=30$ is only about
10\%, which is in contrast to the star lattice where the two non-equivalent NN
bonds differ by a factor of 3.5 \cite{RiSch04}.

\begin{figure}
\centerline{\epsfig{file=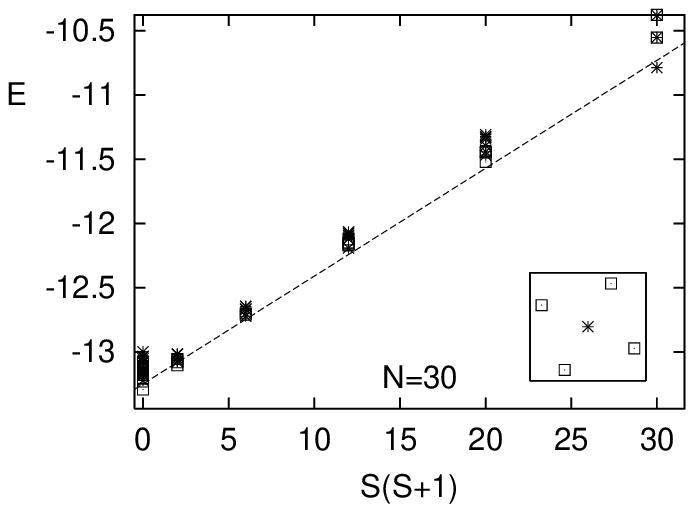,width=0.8\columnwidth}}
\caption{
Low-energy spectrum for the HAFM on the \skagome lattice ($N=30$)
(the inset shows the {\bf k} points in the Brillouin zone).
} 
\label{sqka_spec30}
\end{figure}
Let us now discuss  the 
low-lying spectrum of the star lattice (see Fig.~\ref{sqka_spec30}), 
following the lines of the discussion of the spectrum for the triangular
\cite{bernu94}, the \kagome lattice \cite{lecheminant97,waldtmann98} and the
star lattice \cite{RiSch04}.
The lowest states $E_{\min}(S)$ shown in Fig.~\ref{sqka_spec30} 
are not well described by 
the effective low-energy Hamiltonian 
$H_{\rm eff} \sim E_0 + {\bf S}^2/2N\chi_0$
of a semiclassically ordered system:
One can  see rather clearly
that the dependence $E_{\min}(S)$  vs. $S(S+1)$   is not
a linear one and  
there are no separated so-called 
quasi degenerate joint states \cite{bernu94}
which in the thermodynamic limit could 
collapse to a ground state breaking the
rotational  symmetry. 
Note further that the symmetries of the lowest states in
each sector of $S$ cannot be attributed to the classical ordered ground
states shown in Fig.~\ref{fig2}.
These features are similar to the
\kagome \cite{lecheminant97,waldtmann98} and the star
\cite{RiSch04} lattices. 
But there is one striking difference between  the \kagome
lattice and the star lattice. While the former one has an exponentially
increasing
number  
non-magnetic singlets filling the singlet-triplet gap (spin gap)
 no such low-lying singlets were found for the star lattice
\cite{lecheminant97,waldtmann98,RiSch04}.
This difference was attributed to the non-equivalence of NN bonds in the star
lattice and the resulting dimerization of the ground state.
Though the \skagome lattice has also non-equivalent NN bonds
its spectrum is different from that of the star lattice, rather  it
 shows similar to the \kagome lattice 
a large  number $N_s$ of non-magnetic excitations
within the singlet-triplet gap.  We find 
$N_s=     6$ ($N=12$),   13 ($N=18$),   17 ($N=24$), 47 ($N=30$),  38
($N=36$).
These numbers increase with growing size (except for $N=36$, which might be
attributed to the lower symmetry of this finite lattice)
but are smaller than those for the \kagome lattice \cite{waldtmann98}, where
an exponential increase of $N_s$ with $N$ was suggested. 
Our data for the \skagome lattice do  not allow a
secure conclusion about a possible exponential increasing of $N_s$ with $N$.

For the discussion of magnetic long-range order we use the 
following finite-system order parameter \cite{Richter04,RiSch04}
\begin{equation}
\label{magnetitation}
 m^+=\bigg ( \frac{1}{N^2}\sum_{i,j}
      \left |\langle{\bf S}_{i}{\bf S}_{j}\rangle \right | \bigg)^\frac{1}{2},
\end{equation}
which is independent on any assumption
on  eventual classical order.
The value $m^+_{\rm class}$ for the two ordered classical ground 
states shown in Fig.~\ref{fig2} is
$m^+_{{\rm class}}=
\frac{1}{2}\sqrt{2/3}$, which is 
the same as for the classical \stateX $\;$ and \stateO $\;$
states on the \kagome and on the star lattices.

The numerical values for 
$(m^+)^2$  are collected in Table \ref{tab1}.
The values of $(m^+)^2$ for the \skagome lattice are comparable to those 
for
the \kagome lattice  but are slightly smaller than the
corresponding values for the star lattice \cite{RiSch04}.
\begin{table}
\caption{ \label{tab1}
Ground state energy per bond
$e_0$, ground state degeneracy $d_{GS}$, spin gap $\Delta$ and square of the 
order parameter $(m^+)^2$
of the spin-half HAFM  on finite \skagome 
lattices.
}
\begin{center}
\begin{tabular}{l c c c c c }
 \hline  \hline
 N            &     12    & 18        &  24       & 30        & 36\\ \hline
 $e_0$ ($d_{GS}$)     & -0.226870 (1) & -0.223767 (2)& -0.224165 (1)& -0.221527 (4)& -0.222197 (3)\\
 $\Delta$  &  0.382668 &  0.290191  &  0.263906 &  0.188865 & 0.139550\\
 $(m^+)^2$     &  0.184160 &  0.116455 &  0.086735 &  0.068618 & 0.060475\\
 \hline  \hline
\end{tabular}
\end{center}
\end{table}

To estimate the values of $e_0$, $\Delta$ and $m^+$ for the infinite \skagome 
lattice we have extrapolated the data from Table \ref{tab1} to 
the thermodynamic limit
according to
the standard formulas for the two-dimensional spin-half HAFM
(see, e.g.\ \cite{Richter04,neuberger89,hasenfratz93}), namely
$
e_0(N)=  e_0(\infty) + A_3 N^{-\frac{3}{2}} + {\cal O}(N^{-2})
$
for the ground state energy per bond,
$
m^+(N) = m^+(\infty) + B_1 N^{-\frac{1}{2}} +  {\cal O}(N^{-1})
$
for the order parameter, and
$
\Delta(N) = \Delta(\infty) +  G_2 N^{-1}  + {\cal O}(N^{-\frac{3}{2}})
$
for the spin gap.
In Table \ref{tab2}
the results of these extrapolations
are presented and compared to
those obtained for spin-half HAFM on the \kagome and on the star
lattices.
Our data suggest a small but finite spin gap and a vanishing order
parameter. 

The values of the extrapolated quantities 
of the \skagome lattice are
very close to those of the \kagome lattice. 
Therefore these data clearly yield evidence for a magnetically disordered
quantum paramagnetic ground state of the spin-half HAFM on the \skagome
lattice which is most likely similar to that of the \kagome lattice.

\begin{table}
\caption{ \label{tab2}
Results of the finite-size extrapolation of the ground state energy per bond
$e_0$, the order parameter $m^+$ 
and the spin gap $\Delta$
of the spin-half HAFM on the \skagome lattice. For  comparison we also show 
results for the  
\kagome  and star lattices taken from Ref. \onlinecite{Richter04,RiSch04}.
To see the effect of
quantum fluctuations we present $m^+$ scaled by its classical
value $m^+_{\rm class}$ for the two ordered states shown in Fig.~\ref{fig2}.
(The negative, but very small, extrapolated values for the \skagome and the
\kagome lattices are
an artefact of the limited accuracy of the extrapolation. We interpret
these negative values as vanishing order parameters.)  
}

\begin{center}
\begin{tabular}{l c c c}
 \hline  \hline
 lattice               & \skagome  & \kagome    & star   \\ \hline
 $\ebond$              & $-0.2209$ & $-0.2172$  & $-0.3091$ \\
 $\Delta$              & $0.052$   & $0.040$    & $0.380$  \\
 $m^+/m^+_{\rm class}$ & $-0.032$   & $-0.036$  & $0.122$ \\
 \hline  \hline
\end{tabular}
\end{center}
\end{table}

\section{Magnetization process}
\begin{figure}
\centerline{\epsfig{file=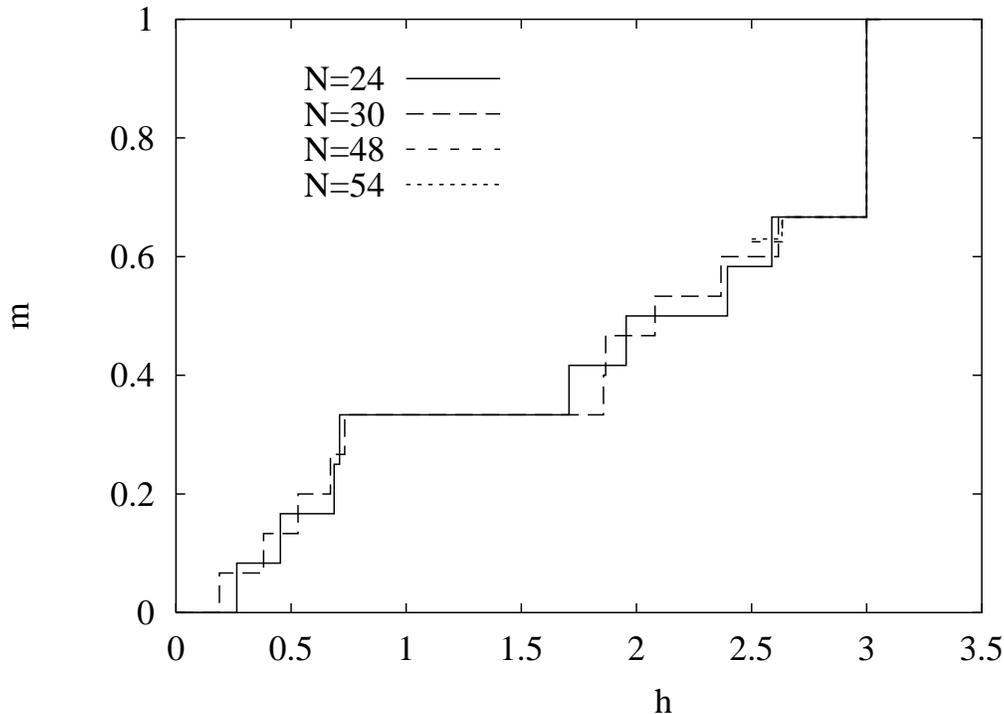,width=0.6\columnwidth, angle=270}}
\caption[]{
Magnetization curves 
of some finite  spin-half HAFM
systems on \skagome lattice in a magnetic field ($N=24,30,48,54$). 
} 
\label{m_h}
\end{figure}
In this Section we briefly discuss 
the magnetization versus field curve 
for some finite
\skagome lattices. 
The magnetization $m$ is defined as
$m= 2 \langle {\hat S}_z \rangle/N$.
We focus on those finite lattices having optimal lattice symmetries, i.e.
$N=24,30$. In the high field sector we are able to present also data for
$N=48$ and $N=54$. 

The results are shown in  Fig.~\ref{m_h}.
Due to the spin gap (see Tables \ref{tab1} and \ref{tab2}) 
one observes a small zero-field plateau.
Clear evidence for a further  plateau is found at $m=1/3$
which can be attributed  to the presence of triangles \cite{hon04}. 
Note that a $m=1/3$ plateau is also observed for the triangular
\cite{NiMi,hon99,hon04,Richter04}, 
the \kagome
\cite{prl02,hida01,cabra02,hon04,Richter04} and the star \cite{Richter04,RiSch04} lattices.

At the saturation field $h_s=3\;$ a jump in the magnetization curve
appears. The presence of this jump was discussed already in Refs.
\onlinecite{schn01,sp04} and is related to the existence 
of independent localized magnon states found for a class of 
strongly frustrated spin lattices \cite{prl02,ri04,Richter04} among them 
the \kagome and the star lattices.
In the case of the \skagome lattice these localized magnons live on the
squares.
The  height of the jump is $\delta m$ and is related to the maximum
number $n_{\max}$ of independent localized magnons which can occupy the lattice.
For the \skagome lattice we have $n_{\max}=N/6$ and consequently 
$\delta m=1/3$. 
We mention, that these localized magnon states are highly degenerate thus 
leading to a finite
residual $T=0$ entropy at the saturation field $h_s=3$
\cite{Richter04,entropy}.
Just below the jump, i.e.\
at  $m=2/3$ there is evidence for another plateau. Its width was estimated
in Ref. \onlinecite{sp04} by finite size extrapolation 
to $\Delta h \approx 0.33J$  for the infinite system.

\section{Summary and conclusions}
In this paper we have discussed the ground state properties of the spin-half
Heisenberg antiferromagnet on the \skagome lattice. This lattice has
similarities with the \kagome as well as with the star lattice. The \kagome
 and the \skagome lattices have coordination number $z=4$ and are built by
corner sharing triangles. The star lattice ($z=3$) shares with the \skagome
lattice the property to have two non-equivalent nearest-neighbor bonds 
and to have an
even number (namely six) sites per unit cell (note that the \kagome lattice
has three sites per unit cell and all nearest-neighbor bonds are 
equivalent). 
On the classical and on the semiclassical level of linear spin wave theory
the ground state of the Heisenberg antiferromagnet on all three 
lattices exhibits very similar properties.
However,  it was argued \cite{RiSch04}
that in the extreme quantum limit $s=1/2$ just these geometrical 
properties of the star 
lattice in common with the \skagome lattice
but different to the \kagome lattice 
lead to different quantum ground states for the star and the \kagome
lattices.
Interestingly, our results for the \skagome lattice lead to the conclusion
that the quantum ground state of the Heisenberg antiferromagnet 
on the \skagome lattice is similar 
to that of the \kagome lattice.
We find evidence for a spin-liquid like ground state with a small gap of
about $J/20$ and a considerable number of low-lying singlets within this
spin gap. Contrary to the star lattice case we do not see here a 
tendency towards forming 
a valence bond crystal ground state.   
The magnetization curve of the $s=1/2$ HAFM on the \skagome lattice
shows a jump just below saturation
and three plateaux at $m=0, \; 1/3$ and $2/3$.
The low-energy excitations present near saturation field 
promise large magnetocaloric effects \cite{entropy}.\\

{\bf Acknowledgment}\\
We thank A. Honecker for stimulating discussions and for critical reading
of the manuscript.
We acknowledge support from the Deutsche Forschungsgemeinschaft
(Projects No.  436 POL 17/3/04 and No. Ri615/12-1)
and from the Polish Committee for Scientific Research
(Project No. 1 PO3B 108 27).


\end{document}